\documentstyle[prl,aps,epsf,floats,twocolumn,graphicx,amssymb]{revtex} 
\begin{document}
\draft
\twocolumn[\hsize\textwidth\columnwidth\hsize\csname
@twocolumnfalse\endcsname

\title{The weighted random graph model}
\author{Diego Garlaschelli}
\address{Dipartimento di Fisica, Universit\`a di Siena, Via Roma 56, 53100 Siena ITALY\\}
\date{\today}
\maketitle
\begin{abstract}
We introduce the weighted random graph (WRG) model, 
which represents the weighted counterpart of the Erdos-Renyi random graph and provides fundamental insights into more complicated weighted networks. 
We find analytically that the WRG is characterized by a geometric weight distribution, a binomial degree distribution and a negative binomial strength distribution. We also characterize exactly the percolation phase transitions associated with edge removal and with the appearance of weighted subgraphs of any order and intensity. We find that even this completely null model displays a percolation behavior similar to what observed in real weighted networks, implying that edge removal cannot be used to detect community structure empirically. By contrast, the analysis of clustering successfully reveals different patterns between the WRG and real networks.
\end{abstract}
]
\narrowtext
The Erdos-Renyi (ER) random graph \cite{bollobas} is the prototype of all unweighted network models: in a graph with $N$ vertices, an unweighted edge is drawn independently between any pair of vertices with equal probability $p$. 
The ER model provides a fundamental reference for the properties of real networks, whose global properties are in most cases expected to arise from surely more complicated, but nonetheless local, decentralized mechanisms. 
Indeed, the explosion of interest towards complex networks \cite{guidosbook} originates precisely because of the striking difference between the observed properties of real networks and the behaviour of the ER model. For this reason, virtually any model of unweighted networks is an extension or modification of the ER random graph. One important example is the class of hidden-variable or fitness network models \cite{fitness,pastor}, where the connection probability is assumed to be no longer a constant $p$ for all pairs of vertices, but a function $p_{ij}=p(x_i,x_j)$, where $x_i$ represents a variable (fitness) associated with vertex $i$. Both the analytical treatment and the intuitive understanding of such more complicated models strongly rely on well-established results for the ER random graph, which thus remains at the basis of whole network theory.

Recently, more and more results indicate that a complete and balanced description of complex networks is only possible if the full degree of heterogeneity present in edge weights is taken into account \cite{vespy_weighted}. Even for purely unweighted graphs, edge weights naturally emerge as dynamical properties, when transport, random walks or other processes take place on the network \cite{transport,vespybook}.
This has triggered a series of studies aimed at generalizing concepts and quantities originally developed for unweighted graphs to weighted networks \cite{vespy_weighted,newman_weighted,intensity,kertesz_clustering,serrano,serrano2,colizza2,myensemble}. 
Quite surprisingly, despite this rapidly evolving context, the studies devoted to the most basic and fundamental graph model - the ER model -   have not been paralleled by analogous efforts to understand weighted networks by comparison with a simple prototype model. 
Indeed, all the weighted network models that have been proposed fall in two broad classes: evolving models \cite{vespy_archi,kertesz_weightedmodel} aimed at reproducing the empirical properties of real networks through generalizations of unweighted growth rules \cite{barabba,guidoassort}, and weighted ensembles obtained by keeping some property of a real network fixed, and randomizing the graph in any other respect \cite{serrano,serrano2,colizza2,mybosefermi}.
Both classes retain some of the heterogeneously distributed properties of real weighted networks, either as the outcome or as the input, and therefore do not represent a completely homogeneous null model, as the ER is for unweighted graphs. 
Therefore, while these models are important for various reasons, they do not allow to decouple two possible sources of heterogeneity.
To see this, note that the properties of interest in real weighted networks are defined as generalizations of their unweighted counterparts \cite{vespy_weighted,newman_weighted,intensity,kertesz_clustering,serrano2,colizza2,myensemble}. 
At present, it is still unclear whether the behaviour observed at the weighted level in real networks arises simply because of the broad range of allowed weight values (a possibility that does not apply for unweighted networks), or because of deeper tendencies and correlations, as empirically found at the unweighted level. 
A weighted counterpart of the ER random graph would exhibit only the first type of heterogeneity, thus representing a novel reference in order to correctly interpret the observed behaviour of weighted networks. 
It would also represent the basis for separately introducing the second type of heterogeneity, since the recently identified correct null models for weighted graphs \cite{mybosefermi}, that allow to preserve features of real networks, can be redefined as sophistications of a homogenous weighted model.

We now introduce a model with the above characteristics, and denote it the \emph{weighted random graph} (WRG).
We follow a straightforward analogy with the unweighted case. 
The ER random graph can be obtained as a particular case of the configuration model, representing the maximally random ensemble of unweighted networks with specified degrees \cite{maslov,newman_origin}. The latter can be described analytically by the connection probability $p_{ij}=x_i x_j/(1+x_i x_j)$, where each parameter $x_i$ allows to control the expected degree $\langle k_i\rangle=\sum_{j\ne i} p_{ij}$ of  vertex $i$. This ensures that all graphs with the same degree sequence $\{k_i\}_{i=1}^N$ are equiprobable \cite{newman_origin}.
If all vertices are assigned the same value $x_i=x_0$, the above probability becomes a constant $p=x_0^2/(1+x_0^2)$ and the ER model (where all vertices have the same expected degree) is recovered. In such a way, all graphs with the same number of links $L$ become equiprobable. 
We define the WRG following an analogous derivation.
In a weighted network, the strength of a vertex $i$ is defined as $s_i\equiv\sum_{j\ne i}w_{ij}$, where $w_{ij}$ is the weight of the edge connecting $i$ and $j$. The weighted analogue of the configuration model is the maximally random ensemble of weighted networks with specified strengths \cite{serrano,serrano2,colizza2,mybosefermi}. 
Recently, it was shown \cite{mybosefermi} that this ensemble can be generated by drawing an edge of weight $w$ between vertices $i$ and $j$ with probability
\begin{equation}\label{eq_qij}
q_{ij}(w)=(y_i y_j)^w (1-y_i y_j)
\end{equation}
where $y_i<1$ allows to tune the expected strength $\langle s_i\rangle$ of vertex $i$.
The case $w=0$ corresponds to no edge being drawn.
The above expression allows $w$ to take discrete, integer values in the range $[0,+\infty)$. If a finite maximum allowed weight $w_{max}$ is considered, the above expression must be divided by $1-(y_i y_j)^{w_{max}+1}$, which however rapidly converges to $1$ as $w_{max}$ increases. In this model, all weighted graphs with the same strength sequence $\{s_i\}_{i=1}^N$ are equiprobable \cite{mybosefermi}. 

In analogy with the ER model, we define the WRG as a weighted graph where all vertices are statistically equivalent, or $y_i=y_0<1$ $\forall i$. Setting $p\equiv y_0^2$ (we will justify this choice in a moment) and assuming for simplicity $w_{max}=+\infty$, we obtain 
\begin{equation}\label{eq_q}
q(w)=p^w (1-p)
\end{equation}
as the probability that any two vertices are joined by an edge of weight $w$. This completely specifies the model. A demonstration allowing to generate small weighted graphs according to the model is available \cite{tiziano_wrg}.

\begin{figure}[]	
\includegraphics[width=.5\textwidth]{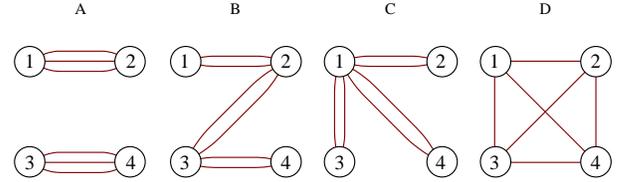}
\caption[]{An incomplete list of equiprobable graphs with $N=4$ vertices in the WRG.}
\label{fig_subgraphs}
\end{figure}

Note that $q(0)$ is the probability that no edge is drawn. Therefore 
$1-q(0)=p$
is the probability of an edge of any (nonzero) weight. If we project the weighted network onto an unweighted graph, we therefore obtain an ER random graph with connection probability $p$, justifying the choice of the symbol. However, for the purpose of fitting the model to a real network, the Maximum Likelihood principle indicates different parameter choices in the two cases. Using the results in ref.\cite{mylikelihood}, for the ER model the optimal parameter choice is $p^*=2L/N(N-1)$ where $L$ is the number of links in the real network, while for the WRG it is easy to show that the likelihood-maximizing choice is
\begin{equation}
p^*=\frac{2W}{N(N-1)+2W}
\end{equation}
where $W\equiv\sum_{i<j}w_{ij}$ is the sum of edge weights in the real network. This is the optimal criterion to be used in order to tune $p$ in the WRG, all other choices resulting in a decreased likelihood \cite{mylikelihood}. In particular, one should not be tempted to use $p$ in order to reproduce the number of links $L$ of the purely topological projection as for the ER model. This reflects the fact that in the WRG model the equiprobable realizations are those with the same total weight $W$, not those with the same number of links $L$.
To illustrate this point, fig.\ref{fig_subgraphs} shows four graphs that are equiprobable in the WRG, since they have the same number of vertices ($N=4$) and the same total weight ($W=6$). The graph A has $l=2$ weighted edges of weight $w=3$, B and C have $l=3$ edges of weight $w=2$, and D has $l=6$ edges of weight $w=1$. Note that the unweighted projections of these graphs have different numbers of links $L$.

The form of $q(w)$ appearing in eq.(\ref{eq_q}) coincides with the so-called geometric distribution with parameter $1-p$. The latter represents the probability that, in a series of independent Bernoulli trials with success probability $p$, $w$ successes occur before a failure occurs for the first time. Thus the WRG is completely equivalent to the following: select one pair of vertices, and start a series of Bernoulli trials with success probability $p$. Each success implies the formation of a link of unit weight between the same two vertices. $w$ multiple successes correspond to a weighted edge of weight $w$.
As soon as a failure occurs for the first time, the sequence of trials stops and a new pair of vertices is selected. The process is repeated until all pairs have been considered. Indeed, this procedure represents a fast and convenient way to generate the WRG my means of numerical simulations \cite{tiziano_wrg}. 
Note that we are considering an undirected weighted model. The extension to the directed case is straightforward.

\begin{figure}[]	
\includegraphics[width=.45\textwidth]{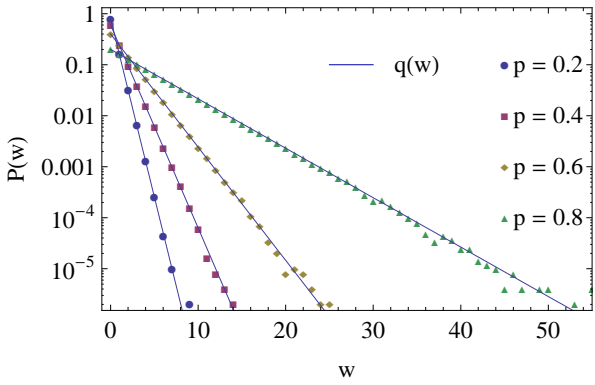}
\caption[]{Weight distribution in the WRG. 
For each choice of the parameter $p$, isolated points correspond to the histogram $P(w)$ of edge weights in numerical simulations of the model with $N=1000$ vertices, and solid lines represent the theoretical curve $q(w)$ obtained for the same value of $p$.}
\label{fig_qw}
\end{figure}

We now derive a series of exact results for the WRG.
Since all weights are independently drawn from $q(w)$, it follows that the weight distribution $P(w)$ is simply
\begin{equation}
P(w)=q(w)
\end{equation}
as confirmed in fig.\ref{fig_qw}. The expected weight of any edge is
\begin{equation}\label{eq_mu}
\langle w\rangle =\sum_{w=0}^{+\infty}q(w)w=\frac{p}{1-p}
\end{equation}
and the variance is
\begin{equation}
\langle w^2\rangle-\langle w\rangle^2 =\frac{p}{(1-p)^2}
\end{equation}
Note that, for networks approaching the fully connected topology ($p\to 1$), the mean and variance of the weight distribution diverge. 

At a purely topological level, the expected degree of any vertex is
$\langle k\rangle =(N-1)p$ as in the ER model, and the probability $P(k_i=k)$ that the degree of vertex $i$ equals the value $k$ follows a binomial distribution with parameters $N-1$ and $p$. The usual approximation for the ER model is to neglect correlations between vertices, which implies that $P(k_i=k)$ coincides with the degree distribution $P(k)$ of the whole network: 
\begin{equation}\label{eq_pk}
P(k)= \left(
\begin{array}{c}
 N-1 \\
 k
\end{array}
\right)p^{k}(1-p)^{N-1-k}
\end{equation}
The mean is $(N-1)p$ and the variance is $(N-1)p(1-p)$.

\begin{figure}[]	
\includegraphics[width=.5\textwidth]{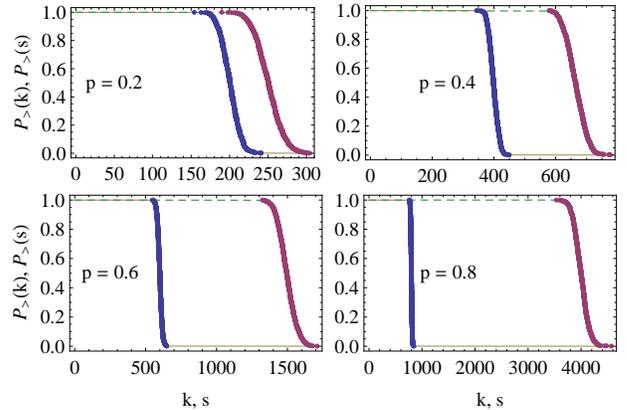}
\caption[]{Cumulative degree and strength distributions in the WRG. 
The four panels correspond to different choices of the parameter $p$. In each panel, points correspond to the cumulative histograms of vertex degrees (left) and strengths (right) in numerical simulations of the model with $N=1000$ vertices, and curves represent the theoretical cumulative distributions $P_>(k)$ (solid) and $P_>(s)$ (dashed).}
\label{fig_pks}
\end{figure}

We now consider the strength distribution $P(s)$. 
The expected strength of any vertex is $\langle s\rangle =(N-1)\langle w\rangle =(N-1)p/(1-p)$.
Using the same approximation leading to eq.(\ref{eq_pk}) (i.e. neglecting correlations between vertices), the strength distribution $P(s)$ of the whole network coincides with the probability $P(s_i=s)$ that $s_i$ equals the particular value $s$. 
Note that $s_i=\sum_{j\ne i} w_{ij}$ is a sum of $N-1$ independent geometrically distributed variables with parameter $1-p$, and its distribution $P(s_i=s)$ is known as the negative binomial distribution with parameters $N-1$ and $1-p$, and reads:
\begin{equation}\label{eq_ps}
P(s)= \left(
\begin{array}{c}
 N-2+s \\
 N-2
\end{array}
\right)p^s(1-p)^{N-1}
\end{equation}
The mean is $(N-1)p/(1-p)$ and the variance is $(N-1)p/(1-p)^2$. The degree and strength distributions are shown in fig.\ref{fig_pks} in cumulative form. Numerical simulations are in perfect accordance with the analytical results. Note that $P(k)\approx P(s)$ when $p\approx 0$, while the two distributions become increasingly different as $p\to 1$: the degrees are distributed with vanishing variance about the average value $\langle k\rangle=N-1$ (fully connected topology), while the strenghts are distributed with diverging variance about a diverging (even for finite $N$) average value.

\begin{figure}[]	
\includegraphics[width=.45\textwidth]{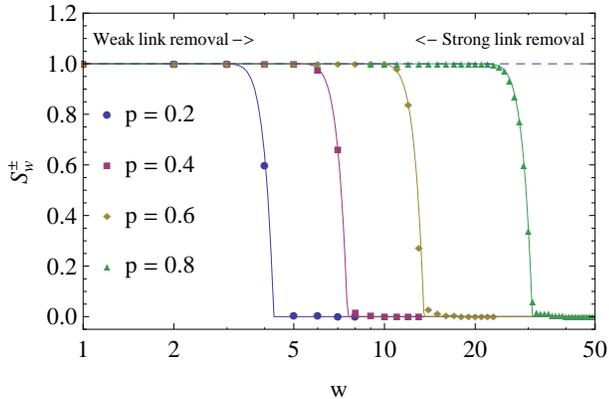}
\caption[]{Fraction of vertices spanned by the LCC after weak ($S^+_w$, solid, left to right) and strong ($S^-_w$, dashed, right to left) link removal. Symbols: numerical simulations with $N=1000$; curves: theoretical results.}
\label{fig_sw1}
\end{figure}

An important property of weighted networks is their resilience against targeted edge removal \cite{kertesz_weightedmodel,removal1,removal2,removal3}.
One is interested in studying the behaviour of the largest connected component (LCC) and the clustering coefficient in the unweighted projection under progressive deletion of edges, in increasing or decreasing weight order. 
In real networks, one observes that the LCC is surprisingly robust under strong link removal, and very fragile under weak link removal  \cite{removal1,removal2,removal3}. In particular, for weak link removal a percolation phase transition where the LCC fragments abruptly at a finite critical weight is observed, while for strong link removal no phase transition is observed. 
This behaviour is often taken as a signature of community structure: the network is interpreted to be organized in communities, with strong intra-community edges and weak inter-community ones \cite{kertesz_weightedmodel,removal1,removal2,removal3}.
We now consider the resilience of the WRG, that can be studied exactly. 
If all edges with weight smaller than $w$ are removed (weak link removal), it is easy to see that the remaining edges form an unweighted projection equivalent to an ER random graph with connection probability
\begin{equation}
p^+_w\equiv\sum_{v=w}^{+\infty}q(v)=p^{w}
\end{equation}
Similarly, if all edges with weight greater than $w$ are removed (strong link removal), the unweighted projection will be an ER random graph with probability
\begin{equation}
p^-_w\equiv\sum_{v=1}^{w}q(v)=p-p^{w+1}
\end{equation}
Thus, in both cases we can exploit a well known result for the ER model \cite{bollobas} to obtain the equation obeyed by the fractions $S^+_w$ and $S^-_w$ of vertices in the LCC after weak or strong link removal (up to the value $w$) respectively:
\begin{equation}
S^{\pm}_w=1-\exp(-z^{\pm}_w S^{\pm}_w)
\end{equation}
where $z^{\pm}_w=(N-1)p^{\pm}_w$ is the average degree after edge removal. 
The theoretical value of $S^{\pm}_w$ is the largest solution to the above equation, which is easily obtained numerically. In fig.\ref{fig_sw1} we show that such solution is perfectly confirmed by numerical simulations.
For weak link removal, as in the ER model a percolation phase transition is observed at the critical probability $p^+_w\simeq 1/N$, corresponding to the critical weight 
\begin{equation}
w^+_c\equiv -\frac{\ln N}{\ln p}
\end{equation}
When edges are removed in increasing weight order, the LCC will span a finite fraction of the $N$ vertices as long as edges with weight $w< w^+_c$ are removed. As soon as edges with weight $w^+_c$ or larger are removed, the network will fragment into many small connected components. 
For a finite network with $0<p<1$, $w^+_c$ is always finite and nonzero, as shown in fig.\ref{fig_sw1}. For infinite networks, $w^+_c$ depends on how $p$ scales with $N$. In particular
\begin{equation}
\lim_{N\to\infty} w^+_c=\frac{1}{\alpha}
\qquad\textrm{if}\quad p\sim N^{-\alpha}
\end{equation}
where $0\le\alpha\le 1$ (note that $\alpha>1$ would imply the absence of the giant component already in the original network).
For strong link removal, the critical probability $p^-_w\simeq 1/N$ defines the critical weight 
\begin{equation}
w^-_c\equiv \frac{\ln (p-1/N)}{\ln p}-1\simeq 0
\end{equation}
which is always zero, both for large but finite networks with $0<p<1$ and for infinite networks, independently of how $p$ scales with $N$.
Therefore we find the surprising result that the strongest links are completely inessential to the robustness of the network: the giant  component is preserved until the weakest edges are removed, and no phase transition is observed for finite $w$. As we mentioned, real networks display exactly such striking difference between weak and strong link removal. However, since this behaviour is displayed even by our completely random model, where communities are clearly absent, we find that the differences between strong and weak link removal are not  a signature of community structure in real networks, in constrast with  usual interpretations \cite{removal1,removal2,removal3}.

\begin{figure}[]	
\includegraphics[width=.45\textwidth]{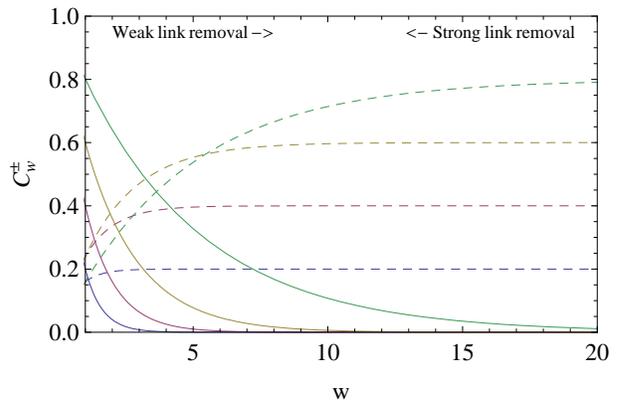}
\caption[]{Average clustering coefficient after weak ($C^{+}_w$, solid, left to right) and strong ($C^{-}_w$, dashed, right to left) link removal. In both cases, four choices of $p$ are considered: from top to bottom, $p=0.8$, $p=0.6$, $p=0.4$, and $p=0.2$.}
\label{fig_cw}
\end{figure}

Opposite considerations apply to the clustering coefficient. After weak or strong link removal, the average clustering coefficient of the unweighted projection $C^{\pm}_w$ simply equals the link density $p^{\pm}_w$. In fig.\ref{fig_cw} we plot $C^{\pm}_w=p^{\pm}_w$ as a function of $w$ for weak (from left to right) and strong (from right to left) link removal. We find a convex, rapidly decaying curve in the former case, and a concave, slowly decreasing curve in the latter case.  Remarkably, this is opposite to what is observed for real weighted networks \cite{removal2,removal3}, where weak link removal results in a concave, slowly decaying curve and strong link removal in a convex, rapidly decaying curve. The striking difference arises because triangles and weights are located uniformly in the WRG, while large-weight triangles are located mainly within communities in real networks. Therefore, unlike the LCC, the clustering coefficient signals community structure successfully.

The above results can be easily generalized to the appearance of weighted subgraphs of any order and intensity. Detecting weighted subgraphs in larger networks is an important and stimulating problem \cite{mitrovic}.
The \emph{intensity} $i(g)$ of a weighted subgraph $g$ has been defined as the geometric mean of its edge weights \cite{intensity}. In an ER random graph with $N$ vertices, a subgraph of $k$ vertices and $l$ links appears almost surely when $p$ scales at least as
$N^{-k/l}$ \cite{bollobas}. This allows to solve the corresponding weighted problems: when does a subgraph with $k$ vertices and $l$ weighted edges, each of weight at least (or at most) $w$, appear almost surely in a WRG of $N$ vertices? 
The edges forming such subgraphs are drawn with probability $p^\pm_w$, respectively. Therefore subgraphs with $k$ vertices and $l$ weighted edges, each of weight at least $w$ or at most $w$ appear almost surely if
\begin{eqnarray}
p^+_w\gtrsim N^{-k/l}\quad &\Rightarrow&\quad p \gtrsim N^{-k/wl}\\
p^-_w\gtrsim N^{-k/l}\quad &\Rightarrow&\quad p \gtrsim N^{-k/l}
\end{eqnarray}
respectively. 
Note that these subgraphs of order $k$ have an intensity $i\ge w$ in the former case and $i\le w$ (with $w>0$) in the latter case.
In particular, one may be interested in the appearance of loops, which in the unweighted case play an important topological role \cite{guidosloops}. Since loops of order $k$ are special subgraphs with $l=k$ edges, weighted loops of intensity $i\ge w$ appear when $p\gtrsim N^{-1/w}$, and of intensity $i\le w$ appear when $p\gtrsim N^{-1}$ independently of $w$ (if $w>0$). In both cases, loops of any order and with the same intensity appear simultaneously, as in the unweighted case. The former condition confirms that, in order to have a giant component (which is characterized by loops of all orders) made of edges of weight at least $w$, one must have $p\gtrsim N^{-1/w}$. Similarly, the appearance of weighted loops confirms the behaviour under edge removal of the clustering coefficient $C^{\pm}_w$, which is contributed by triangles (loops of order 3).

We have introduced the WRG as the weighted conterpart of the ER random graph, and derived many of its properties exactly.
The WRG displays the weighted properties that are merely due to the intrinsic variability in edge weights, and not to true correlations. Therefore it is a fundamental reference for the analysis and interpretation of the properties of real weighted networks. Sophistications of the model, in particular allowing different edges to be governed by different parameters as in eq.(\ref{eq_qij}), allow to extend the model to power-law distributed strengths or degrees \cite{mybosefermi}.

\end{document}